\documentclass{aa}
\usepackage{graphicx}
\begin{document}
\newcommand{\kms}{{\, \rm km~s}^{-1}}
\newcommand{\ho}{{\rm km~s}^{-1}~{\rm Mpc}^{-1}}
\newcommand{\obj}{HE~1503+0228}
\title{On-axis spatially resolved spectroscopy of low redshift quasar host 
galaxies: \obj, at $z=0.135$
\thanks{based on observations made with ANTU/UT1 at 
ESO-Paranal observatory in Chile (program 65.P-0361(A)), 
and with the ESO 3.5m NTT, at La Silla observatory 
(program 62.P-0643(B)).}}

\author{F. Courbin \inst{1, 2}
\and G. Letawe \inst{1}
\and P. Magain \inst{1}
\and L. Wisotzki \inst{3}
\and P. Jablonka \inst{4}
\and K. Jahnke \inst{5}
\and B. Kuhlbrodt \inst{5}
\and D. Alloin \inst{6}
\and G. Meylan \inst{7}
\and D. Minniti \inst{2}
\and I. Burud \inst{7}}

\offprints{G. Letawe}
\institute{Institut d'Astrophysique et de G\' eophysique, Universit\' e de
  Li\`ege, All\'ee du 6 Ao\^ut, 17, Sart Tilman (Bat. B5C), Li\`ege 1, Belgium
\and Universidad Cat\'olica de Chile, Departamento de 
Astronomia y Astrofisica,
Casilla 306, Santiago 22, Chile
\and
Institut f\"ur Physik, Universit\"at Potsdam, Am Neuen Palais 10, 
14469 Potsdam, Germany
\and
GEPI, Observatoire de Paris, Place Jules Janssen, F-92915 Meudon Cedex, France
\and
Hamburger Sternwarte, Universitaet
Hamburg, Gojenbergsweg 112, D-21029 Hamburg, Germany
\and 
European Southern Observatory, Casilla 19, Santiago, Chile
\and Space Telescope Science institute, 3700 San Martin Drive, 
Baltimore, MD 21218 USA
}

\date{}

\abstract{We present the first result of a comprehensive spectroscopic
study of quasar host galaxies.  On-axis, spatially resolved spectra of
low redshift  quasars have  been obtained with  FORS1, mounted  on the
8.2m ESO  Very Large Telescope,  Antu. The spectra are  {\it spatially
deconvolved} using a spectroscopic  version of the ``MCS deconvolution
algorithm''. The algorithm decomposes two dimensional spectra into the
individual spectra of  the central point-like nucleus and  of its host
galaxy.  Applied  to \obj\,  at  $z=0.135$ (M$_B$=$-$23.0),  
it  provides us  with  the
spectrum  of   the  host   galaxy  between  3600\AA\,   and  8500\AA\,
(rest-frame), at a  mean resolving power of 700. The  data allow us to
measure several of the important Lick indices. The stellar populations
and gas ionization state of the host galaxy of \obj\, are very similar
to  the   ones  measured  for  normal   non-AGN  galaxies.   Dynamical
information is  also available for  the gas and stellar  components of
the  galaxy.   Using   deconvolution  and  a  deprojection  algorithm,
velocity curves are derived for  emission lines, from the center up to
4\arcsec\,  away   from  the nucleus of the galaxy.    
Fitting  a  simple
three-components mass model (point  mass, spherical halo of 
dark matter, disk) to the
position-velocity  diagram, we infer  a mass  of M(r$<$1kpc) = $(2.0 \pm 0.3) 
10^{10}$ M$_{\odot}$ within the central kiloparsec of the galaxy, and a 
mass  integrated  over  10  kpc  of  M(r$<$10kpc) = $ (1.9  \pm 0.3)  10^{11}$
M$_{\odot}$, with an additional 10\% error due to the uncertainty on the 
inclination of the galaxy.  This, in combination  with the  analysis of  
the stellar
populations  indicates that  the host  galaxy  of \obj\,  is a  normal
spiral galaxy.   
\keywords{Galaxies: dynamics, stellar populations --
quasars: individual: HE~1503+0228 -- techniques: deconvolution}}

\titlerunning{on-axis spectroscopy of quasar hosts I: HE~1503+0228}

\maketitle

\begin{figure*}[t]
\centering
\includegraphics[height=7.8cm]{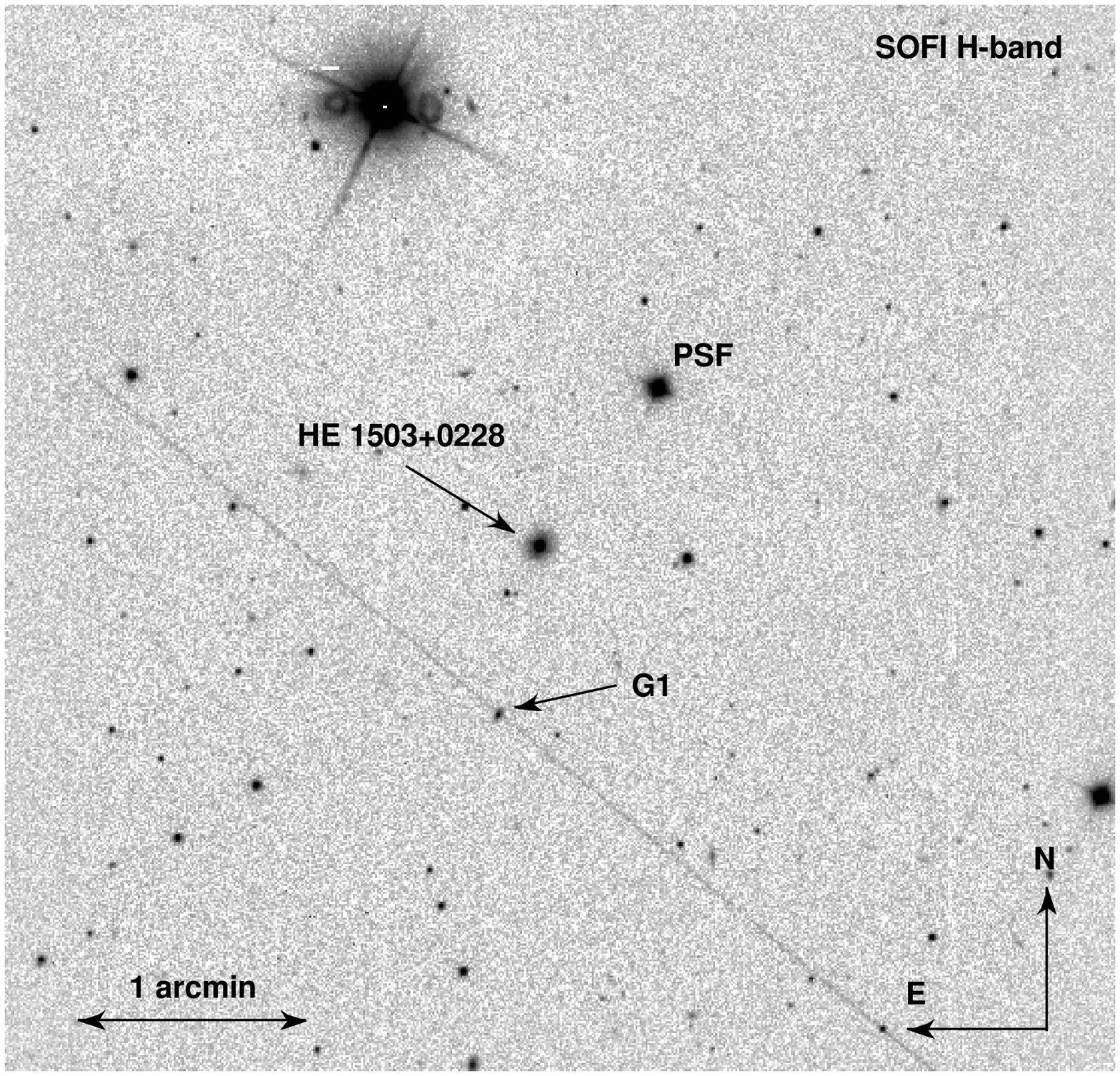}
\leavevmode
\includegraphics[height=7.8cm]{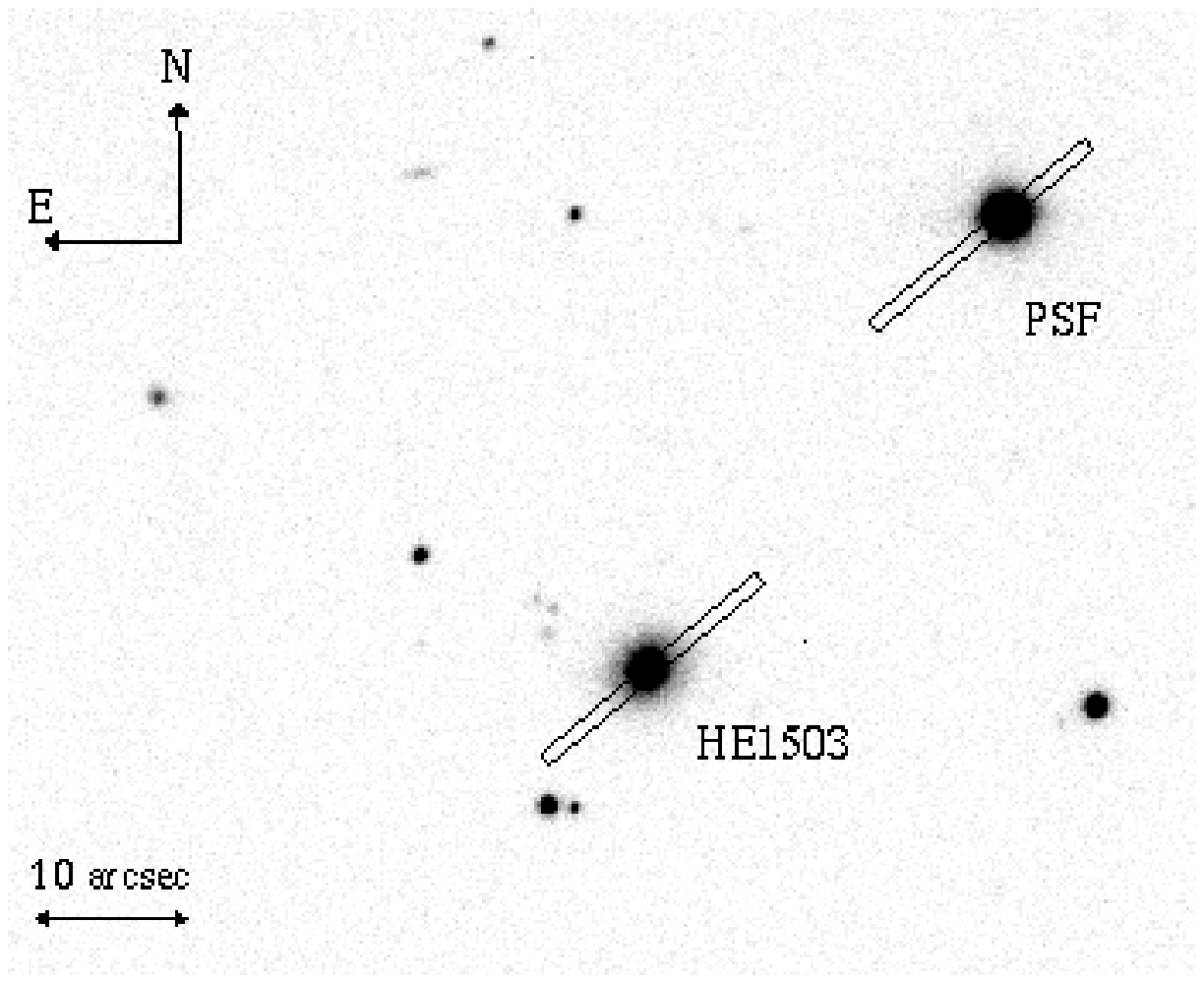}
\caption{{\it  Left:} field  of view  of 5\arcmin  $\times$  5 \arcmin
around  \obj. The exposure  ($H$-band) is  480 sec  long. It  has been
obtained with  SOFI on  the ESO 3.5m  NTT. The seeing  is 0.8\arcsec\,
sampled with a 0.29\arcsec\, pixel. {\it Right:} part of the VLT/FORS1
pointing  image. The  seeing  is 0.62\arcsec  on  this 30sec  $R$-band
exposure.  The 1\arcsec\,  slits used  to  obtain the  spectra of  the
target and of the PSF star are indicated.}
\label{field} 
\end{figure*}
\begin{figure*}
\centering
\includegraphics[height=5.7cm]{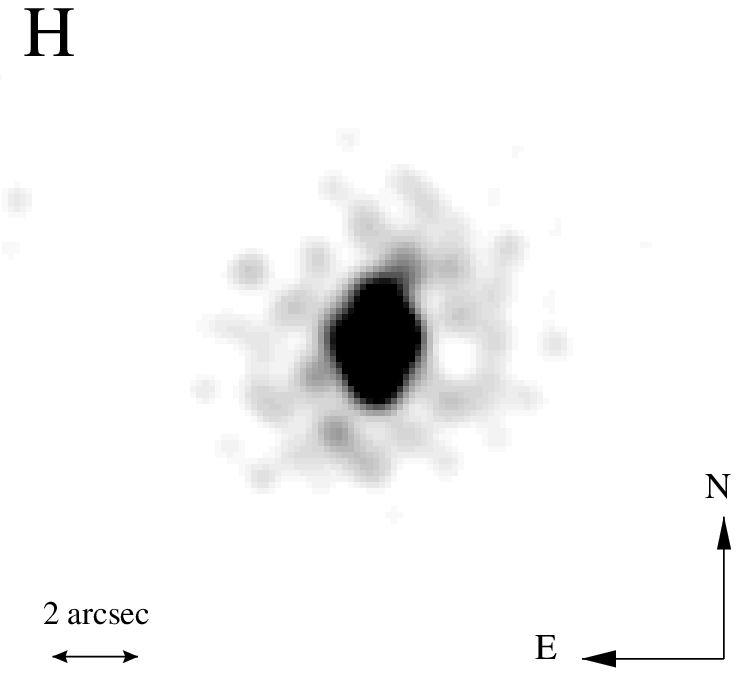}
\leavevmode
\includegraphics[height=5.7cm]{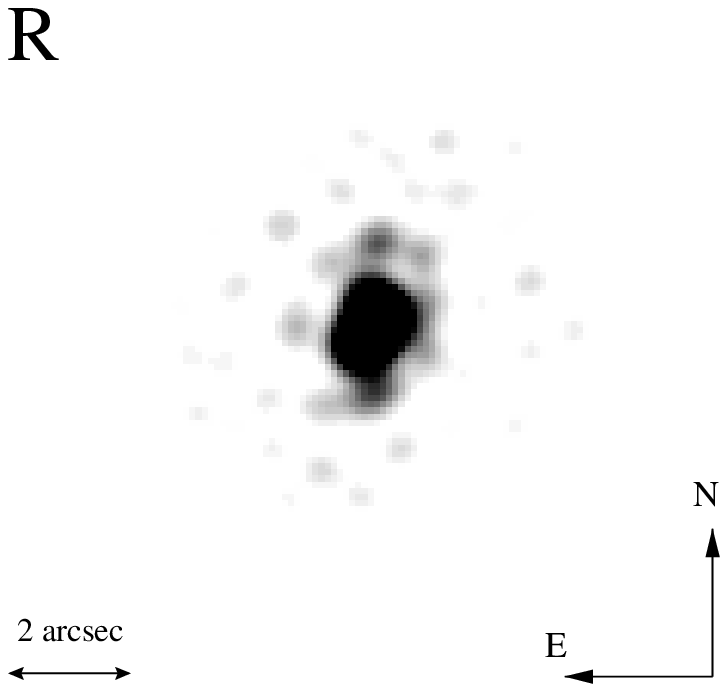}
\leavevmode
\includegraphics[height=5.7cm]{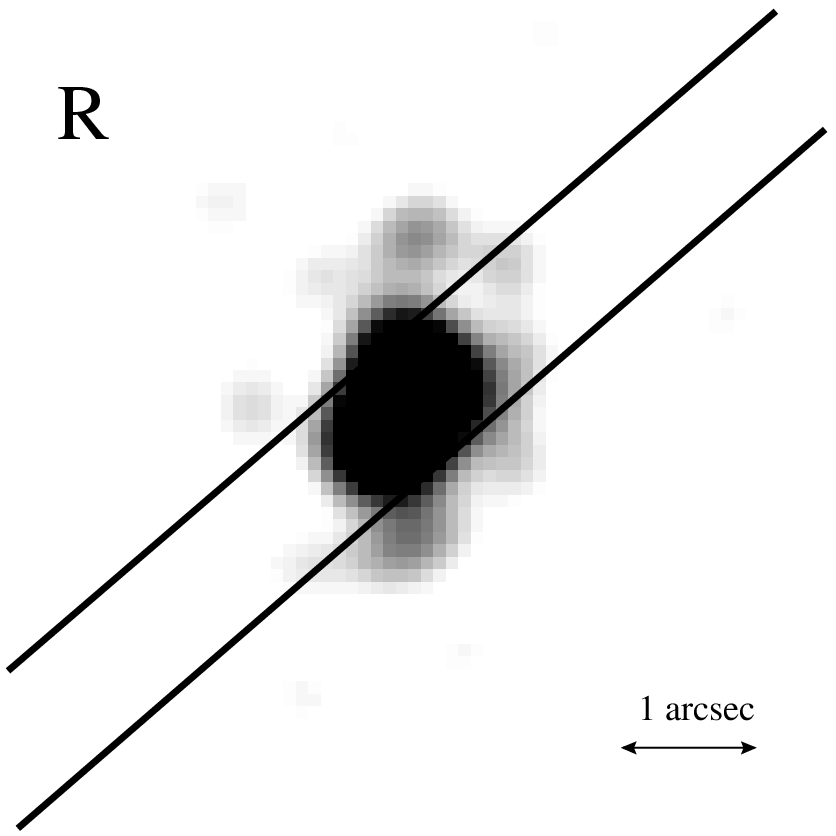}
\caption{{\it Left and middle:}  deconvolved NTT and VLT images, where
the quasar  has been removed from  the data. {\it Right:}  zoom on the
host  galaxy  and   position  of  the  slitlet  used   to  obtain  the
spectroscopic data.}
\label{deconv} 
\end{figure*}

\section{Introduction}

Luminous quasars are now known to be generally located at the cores of
large  galaxies.  With  the recent  discovery that  supermassive black
holes appear  to reside in most  if not all  galaxies with substantial
bulges (e.g., Magorrian et al.  \cite{Mago98}), it is now quite likely
that  quasar-like  activity  is  an  extremely  common  but  transient
phenomenon,  linking the  growth of  now  extinct black  holes to  the
violent processes during phases  of active accretion (cf.  McLeod et
al. \cite{leod99}).   However, very  little of the  physical processes
operating  during  such  episodes  of  nuclear  activity  is  actually
understood: neither do  we know the actual conditions  for fueling (or
refueling)  massive  black holes,  nor  the  time-scales involved.   A
particularly  interesting problem  is the  importance of  the feedback
from  a high-luminosity  Active Galactic  Nucleus (AGN)  onto  its host
galaxy, in the  form of huge quantities of  ionizing radiation as well
as possible mechanical outflows (jets).

Quasar host  galaxies have been studied almost  exclusively by imaging
(e.g.,   Bahcall   et   al.    \cite{bahcall97},   Stockton   et   al.
\cite{Stockton98}, M\'arquez  et al \cite{Marquez01}).   Recent Hubble
Space Telescope optical  studies have established that high-luminosity
quasars  generally reside  in big  ellipticals, irrespective  of radio
properties   (McLeod   \&   Rieke   \cite{leod95},   Disney   et   al.
\cite{disney95}, Hughes et al.  \cite{hugh00}).  There also seems to
be a trend that more luminous QSOs are hosted by more massive galaxies
(McLure et al.  \cite{Lure99}, McLeod \& Rieke \cite{leod95}).

A  more detailed  understanding of  the physical  conditions  in host
galaxies     can    only     be     obtained    from     spectroscopic
observations. Available  evidence in  this field is  extremely scarce,
and although extensive quasar  host spectroscopy was conducted already
in the early 1980s (e.g.,  Boroson et al. \cite{boroson85}), they have
never  really  been  followed  up with  improved  instrumentation  and
analysis  techniques, except for  a few  isolated objects  (among them
3C~48  being the  probably best-studied  case in  this field  -- e.g.,
Chatzichristou   et   al.     \cite{cha99},   Canalizo   \&   Stockton
\cite{canal2000}). Particularly important is  the fact that nearly all
these  observations  up  to   now  were  designed  as  ``off-nuclear''
spectroscopy, avoiding  the strong contamination from  the AGN itself,
but  yielding only  information  on  the host  for  distances of  $\ga
5-10$\,kpc from the nucleus.

\begin{figure*}
\centering                   
\includegraphics[width=8.7cm]{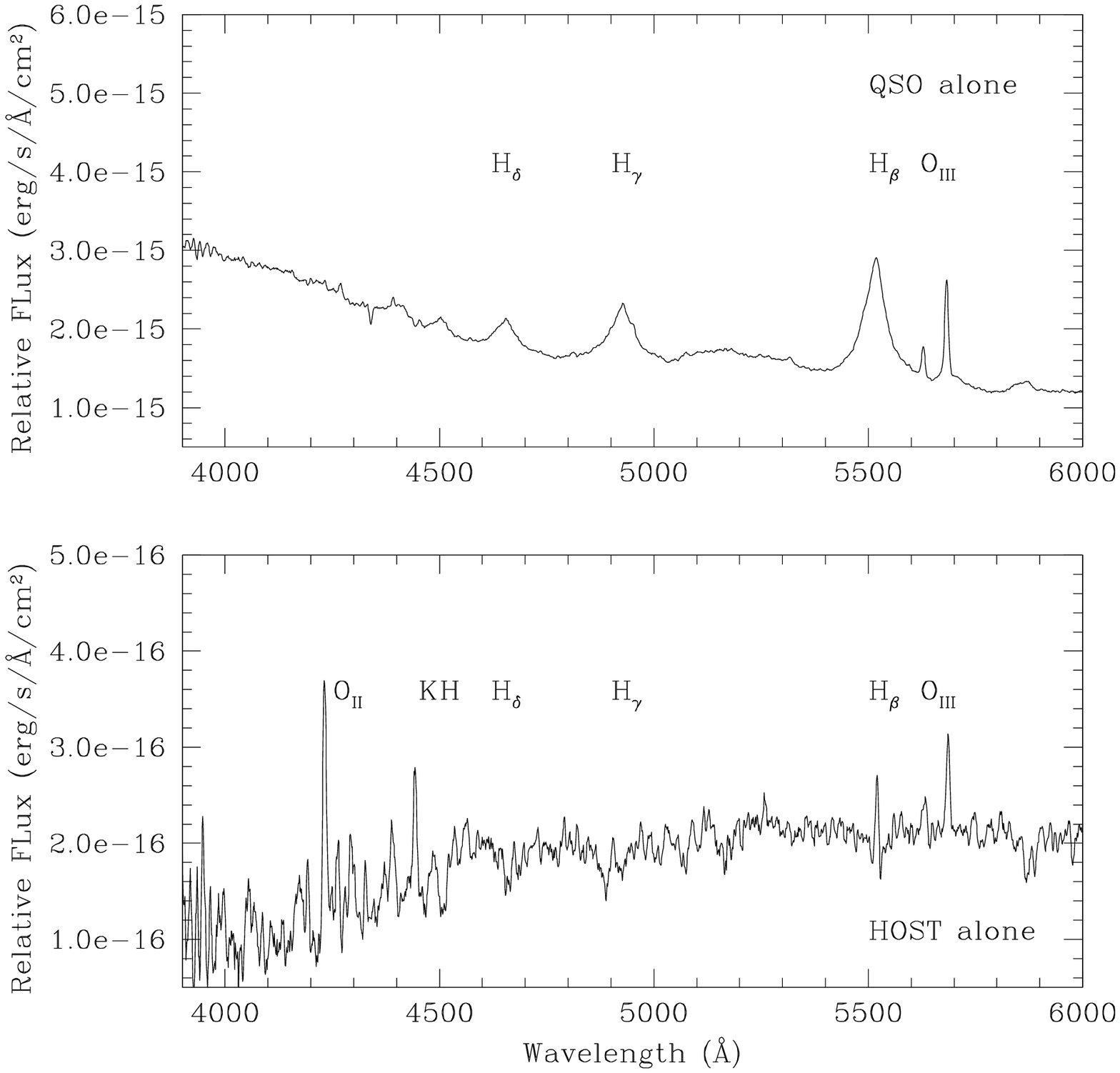}
\includegraphics[width=8.7cm]{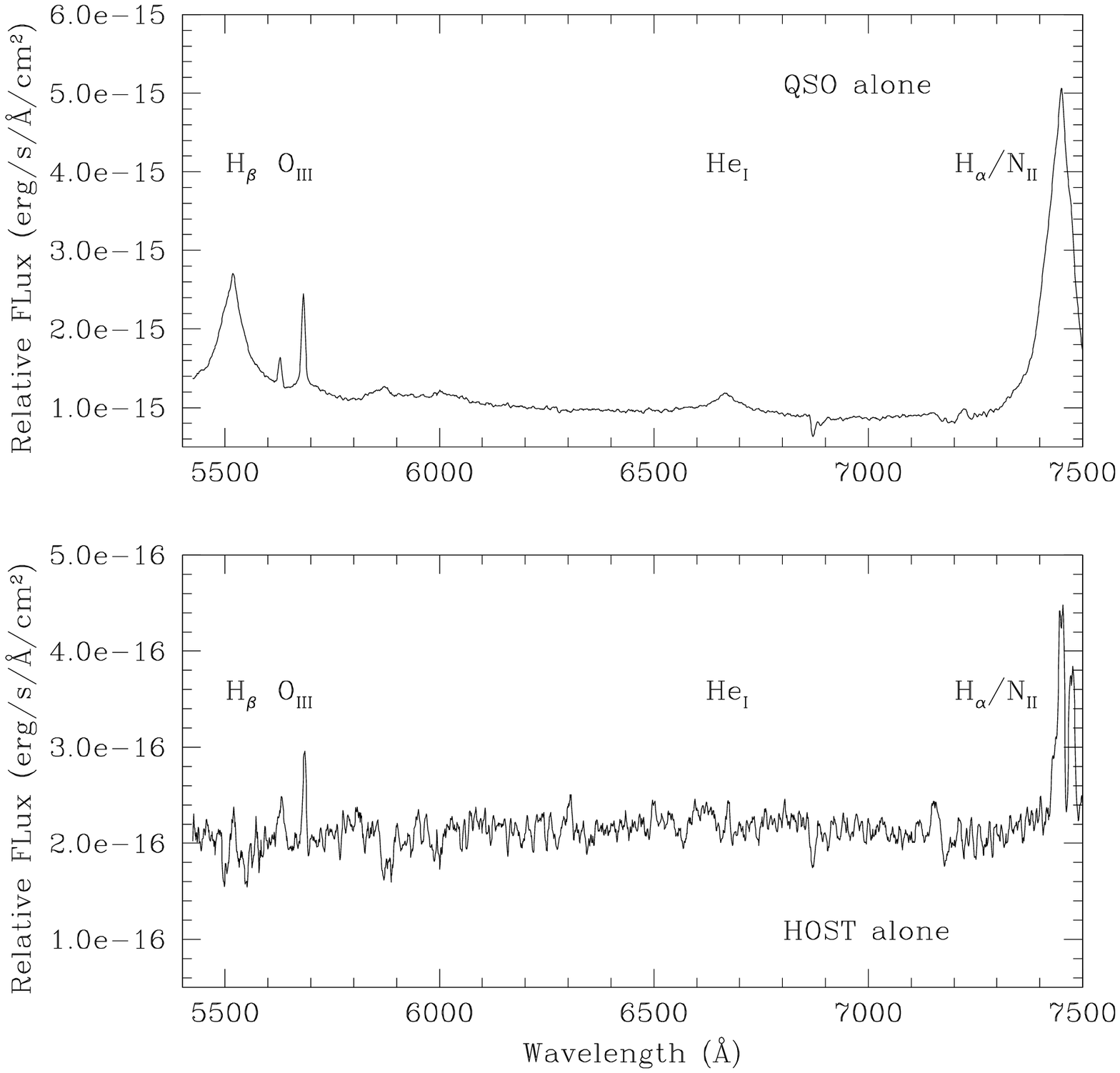}
\includegraphics[width=8.7cm]{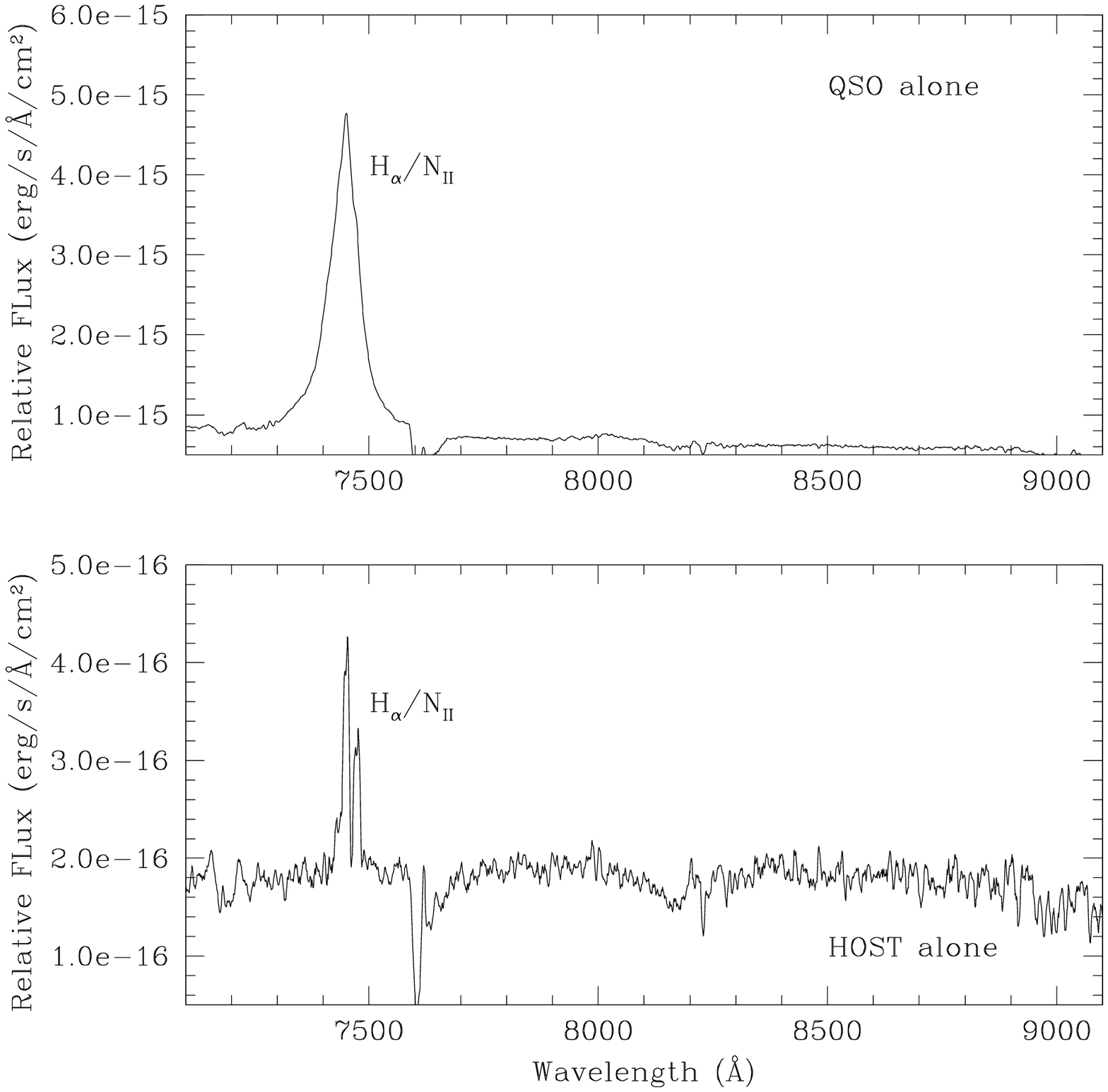}
\caption{One dimensional  deconvolved spectrum of \obj\,  and its host
galaxy. Each  panel corresponds  to one FORS1  grism and  displays the
individual flux calibrated spectrum of  the quasar and its host alone.
Note that the  host spectrum shows no trace of  the AGN broad emission
lines, that  are seen narrow  in the host  galaxy. Note also  the very
good agreement in the deconvolution of the 3 grisms in the overlapping
regions.}
\label{onedspec} 
\end{figure*}

With the  aim of studying the  stellar and gas content  of quasar host
galaxies, as well as their dynamics, we have initiated a spectroscopic
campaign with the ESO-VLT to  observe a sample of luminous radio-quiet
quasars  at low  redshift. The  sample  is based  on the  Hamburg/ESO
Survey (HES; Wisotzki et  al.\ \cite{wisotzki00}), a wide-angle ($\sim
7500$~deg$^2$) search  for optically bright ($B \la  17.5$) quasars in
the  southern sky.   This  survey is  particularly  suited to  provide
samples for  host galaxy studies,  mainly because of two  reasons: (1)
because  of   its  bright   limiting  magnitude,  the   survey  yields
low-redshift  quasars in  large numbers,  (2) the  selection  does not
encompass  morphological criteria,  i.e.,  there is  no limitation  to
point sources as  in most other optical quasar  surveys. This fact and
the wide range of selection criteria employed in the HES (cf. Wisotzki
et  al.  2000)  ensure  that  quasars  are  selected  irrespective  of
morphological properties of their  host galaxies (see also K\"ohler et
al.\ \cite{koehler97}). Our objects all  have $M_B < -23$ and  $z<0.33$.


The present paper  is aimed at exposing the  techniques used to obtain
high  quality  spectra  of   quasar  hosts,  decontaminated  from  the
contamination  by the  quasar, and  to show  their application  to one
object, taken as a test case: \obj, at $z=0.135$. 

\section{Observations -- data reduction}

\subsection{Imaging}

We have obtained $H$-band images of \obj, at La Silla observatory
(ESO, Chile) on  the night of 1999 February  24$^{th}$. SOFI was used,
the near-IR  imager of  the 3.5m ESO  New Technology  Telescope (NTT).
The pixel size  was 0.29\arcsec\, and numerous dithered  images of the
field were taken  in order to produce a  combined sky subtracted frame
with  a total exposure  time of  480 sec  and a  final mean  seeing of
0.8\arcsec.   The  image  is  displayed  on the  left  panel  of  Fig.
\ref{field}.  It  shows the  environment of the  quasar on  a 5\arcmin
$\times$ 5\arcmin\, field, with no obvious companion.

Additional  data  were  obtained  with  the first  of  the  four  8.2m
telescopes (UT1) at Paranal  observatory. A shallow 30s $R$-band image
was taken, in  order to design the spectroscopic  mask. The seeing was
0.62\arcsec, sampled with a pixel scale of 0.2\arcsec.

Both the  near-IR and  optical images were  deconvolved using  the MCS
deconvolution algorithm (Magain et al. 1998). While one single optical
image of \obj\,  was obtained, many dithered IR  images were available
in the $H$-band.   Eight stacks of images were  produced with the data
and   simultaneously   deconvolved   as   explained  in   Courbin   et
al. (1998). This procedure allows for better decomposition of the data
into a sum of point source and ``background'' channels, hence a better
extraction  of the  image of  the host, uncontaminated  by the
AGN's  light. The  deconvolved data  are shown  in  Fig. \ref{deconv},
where the quasar has been subtracted.

The  photometry  of the  host  of \obj\,  has  been  performed on  the
deconvolved images, in a 10\arcsec\, diameter aperture. The magnitudes
we infer in that way are $R=16.78 \pm 0.20$ and $H=14.57 \pm 0.20$.

Determining  the  shape  of  the   galaxy  is  of  importance  as  the
inclination  of  the  disk  component  in the  mass  model  (see  next
sections) depends  on the observed  galaxy shape, hence  affecting the
mass determination  of the  galaxy.  We make  the assumption  that the
galaxy  has  an  intrinsic  circular  geometry,  to  turn  the  fitted
ellipticities into an inclination angle.  Given the relatively low
signal-to-noise of our  images we chose to determine  the shape of the
host by fitting analytical profiles  to the data (Kuhlbrodt et al.\ in
preparation).  Our model  is  composed of  a  nucleus, a  disk, and  a
spheroid,  convolved  with the  observed  PSF  before  any fitting  is
performed.  In the  case of  \obj\, the  spheroid is  used to  model a
bar-like feature in the center of  the galaxy, that may affect the fit
of the disk component if it were not taken properly into account.

The analysis  yields the Position  Angle (PA=$38\degr\pm3\degr$) of
the  host, its  inclination  ($i=32\degr\pm2\degr$),  and the  angle
($\phi=12\pm3\degr$) between the  slit and the  major axis of  the 
galaxy. These values  are in fact very close to  the values found just
by measuring the outmost isophotes in the VLT images.

\subsection{Spectroscopy}
\label{obsspec}

The  spectroscopic  data  were  taken  on the  night  of  2000,  April
11$^{th}$ with FORS1  mounted on Antu, the first of  the four 8.2m VLT
Unit Telescopes at Paranal observatory (ESO, Chile). Three grisms were
used (G600B, G600R,  G600I) to cover the whole  optical spectral range
4000-9500\AA\,  with a  mean spectral  resolution of  about  700.  The
seeing during the observations  was good, about 0.6\arcsec. Each grism
was exposed 1200 sec, under  photometric conditions and dark time,  
which yields a mean S/N of about 150-200 per pixel (quasar+host). The
large pixel size of the detector was used, 0.2\arcsec.

We  show  in  the  right  panel  of  Fig.   \ref{field}  and  in  Fig.
\ref{deconv} the  orientation of the  slitlets relative to  the quasar
host and  to the PSF star.  In Fig. \ref{deconv}, the images  have 
been deconvolved  and the quasar
has been removed from the data.  The FORS mask consists of 19 slitlets
with  fixed length  and width  (19\arcsec x  1\arcsec).  One  slit was
centered on \obj,  one slit was centered on a bright  PSF star and the
rest  of  the  slits were  used  either  to  observe fainter  PSF  or
galaxies.   Galaxy G1  is indicated  in Fig.   \ref{field}.  It  is at
$z\sim 0.3$ and is unrelated to the quasar.

At the  adopted spectral  resolution, the area  accessible on  the CCD
without  loosing spectral  information at  the edges  of the  array is
restricted to  a strip of  sky about 40\arcsec\,  wide. As we  have to
observe simultaneously the quasar and at least one PSF star, the PA of
the mask is therefore imposed  to us by the instrumental design, i.e.,
the 40\arcsec\,  strip of sky has  to include both the  quasar and the
PSF star(s).   This means that the  slitlet centered on  the quasar is
not  necessarily oriented along  one of  the axes  of symmetry  of our
targets.   It  also  means  that  we do  not  observe  at  parallactic
angle. This is  however not a major problem as  FORS1 has an efficient
atmospheric  refraction corrector  and the  data remain  unaffected by
slit losses.

A two-dimensional  wavelength calibration  was applied  in order to
correct for slit curvature, followed  by a
two-dimensional sky  subtraction.  The spectrum  of \obj\, and  of the
PSF star were reduced exactly in  the same way, and they were rebinned
to a common pixel size of 1\AA\, in the spectral direction.  They were
also rebinned to  a common starting wavelength, which  ensures that to
each spectral  resolution element of the quasar,  corresponds the same
spectral resolution element in the PSF.

\section{Extraction of the host's spectrum}

Most current  spectroscopic  studies  of  quasar 
host  galaxies  are  designed  as
off-axis observations, with the slit of the spectrograph placed one to
several times the size of the  seeing disk away from the center of the
AGN (e.g., Hughes et al. \cite{hugh00}).

While such  an observational  strategy is minimizing  contamination by
the quasar,  it has  severe drawbacks. First,  one can only  study the
outer parts  of the galaxy. This  is limiting the  study to particular
regions of  the hosts.  It also  requires very long  exposure times on
large  telescopes   to  detect  the  extremely  faint   wings  of  the
hosts. Second,  all velocity  information is lost,  as is  the spatial
information in general, along  the galaxy's axes of symmetry. Finally,
the method is  still restricted to the most  extended objects and does
not allow a full decontamination of the quasar's light.

Instead, our observations  are taken on-axis, placing the  slit of the
spectrograph  on  the  quasar's  center  of  light.  This  avoids  all
drawbacks   of   the  off-axis   observations,   but  requires   clean
post-processing techniques in order  to remove accurately the spectrum
of the AGN.  For this purpose  we use the spectroscopic version of the
MCS deconvolution algorithm (Magain  et al.  \cite{Magain}, Courbin et
al. \cite{Courbin_a}).  Using the {\it  spatial} information contained
in the spectrum of PSF stars  located close to the object of interest,
the algorithm sharpens the  data in the spatial direction, sub-samples
them in order to achieve  smaller pixel size, and decomposes them into
two channels  containing ({\bf i}) the  spectrum of the  AGN and ({\bf
ii}) the two-dimensional spectrum of the host galaxy.

In  the  deconvolved  spectrum,  the  new sub-sampled  pixel  size  is
0.1\arcsec\,   and  1\AA\,   in  the   spectral  direction.   In  Fig.
\ref{onedspec}  are  displayed  the  one dimensional  flux  calibrated
spectra,  integrated along  the spatial  direction on  the deconvolved
data. Interestingly,  the three FORS1 grisms  have significant regions
of  overlap  that are  in  perfect  agreement  with one  another.   In
addition, at the redshift of  \obj, most emission lines are present in
two  grisms.  This  allowed  us  to  construct  a  combined  2-D  flux
calibrated spectrum.  All measurements presented in this paper, except
for velocity curves, are done on the final combined spectrum.

\section{Dynamics of the host galaxy}

\subsection{Redshift}

From the different emission lines of  the quasar and of the galaxy, we
can extract a sharp value for the redshift of \obj, by simple Gaussian
fitting. The lines of the host galaxy in the deconvolved spectrum have
to be  measured with particular care:  in the center of  the host, the
lines are double,  due to the velocity field of the  galaxy and to the
deconvolution  method. This is mainly due to the extreme contrast
between the narrow emission lines of the host, and the broad lines of
the quasar: some of the flux in the sharpest parts of the host's
emission lines is taken in the point-source component (the quasar)
of the deconvolved spectrum, i.e., the central parts of the lines
are dimmed. We  have  therefore taken  the mean  wavelength
between the  two sides  of emission lines,  when measuring  the host's
redshift.   The   results  for   each   line   are   given  in   table
\ref{redsh}.  The   details  of  the  individual   emission  lines  is
accessible via  a method  described in Section  \ref{velcurves}, which
does not alter the center of the lines.

\begin{table}[t]
\centering 
\caption[]{Redshift of \obj, as infered from the emission lines of the
host galaxy alone, and from the quasar alone. The redshifts are the same
for the quasar and for its host. }
\begin{tabular}{lccc}
\hline
Line & Redshift    & Redshift          \\ 
     & (host only) & (quasar only)   \\
\hline\hline
O\ion{II}   (3727\AA)  & 0.1353 & 0.1352     \\
H$\beta$    (4861\AA)  & 0.1357 & 0.1352  \\
O\ion{III}  (4959\AA)  & 0.1362 & 0.1349    \\
O\ion{III}  (5007\AA)  & 0.1356 & 0.1348   \\
O\ion{I}    (6300\AA)  & 0.1355 & 0.1352     \\
N\ion{II}   (6548\AA)  & 0.1352 &  $-$      \\
H$\alpha$   (6563\AA)  & 0.1353 & 0.1352 \\
N\ion{II}   (6583\AA)  & 0.1354 &  $-$    \\
S\ion{II}   (6716\AA)  & 0.1356 & 0.1343       \\
\hline
Mean                   & $<$0.1355$>$ & $<$0.1350$>$   \\
Mean error             & ($\pm$0.0003) & ($\pm$0.0003)  \\
\hline
\end{tabular}
\label{redsh}
\end{table}

\subsection{Velocity curves: extraction method }\label{velcurves}

In   the  spectroscopic   version   of  the   MCS   algorithm  (Courbin   et
al.  \cite{Courbin_a}), the difference  in spatial  properties between
the AGN  (point source)  and the host  galaxy (extended  component) is
used to separate the spectra of  these two sources. A smoothing in the
spatial  direction  is  applied  for  this  purpose  to  the  extended
component. While  such a decomposition  method is well suited  to most
spectral regions of the host,  it is not optimal for deriving velocity
curves: spatial smoothing  (which is not precisely known  as it varies
with the local S/N) may modify the spectral position of the lines at a
given spatial  position, by averaging spatial  components of different
radial velocities.

We have devised  another method which is better  suited for extracting
radial  velocity curves  of the  host  galaxy.  Instead  of using  the
difference in {\it spatial} properties to extract the spectra, we take
advantage of the  different {\it spectral } properties  of the AGN and
host to determine the radial velocity as a function of distance.

The fact  that the spectral lines  of the host are  much narrower than
any  feature in the  AGN spectrum  is used  to determine  the spectral
position and intensities  of the host spectral lines  as a function of
spatial position.  More  specifically, at each spatial coordinate
we fit the  sum of a Gaussian profile  of variable width representing
the host emission line, plus  a smooth numerical curve accounting for
the AGN spectrum plus the continuum of the host. 

\begin{figure}[t]
\includegraphics[width=8.3 cm]{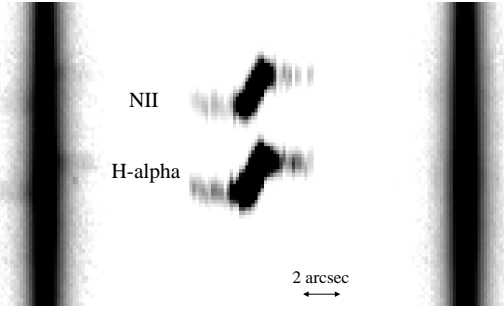}
\caption{Example   of  emission   lines  extraction.   Here   are  the
H${\alpha}$ and [N\ion{II}]  \ emission lines in $I$-band grism. The left part
shows a  zoom on  initial reduced spectrum.  The middle one  shows the
emission lines alone whereas the right part shows the quasar and galactic
continuum.}
\label{ss}
\end{figure}

In doing  this fit, we  also take advantage  of the constant  shift in
wavelength  between two  given emission  lines:  several emission
lines  are fitted simultaneously  under the  constraint that  they are
separated by  a fixed gap  in wavelength. Thus,  when we refer  to the
H${\alpha}$  velocity  curve,  we  in  fact use  the  H${\alpha}$  and
[N\ion{II}] emission lines  simultaneously. Similarly, the H${\beta}$
curve  is  obtained from  the  H${\beta}$ and the  [O\ion{III}]
(5007\AA)  lines, simultaneously . An  example of  such  an extraction  
is displayed  on Fig. \ref{ss}.

In  converting the  wavelength shifts  into velocity  curves,  we take
several effects into account:

\begin{enumerate}

\item[{(}1{)}] The orientation of the  slit does not coincide with the
major axis  of the galaxy.  This is due to  observational constraints,
i.e.  the fact  that  we had  to obtain  the  spectrum of  a PSF  star
simultaneously with the quasar and host (see Section 2.2).

\item[{(}2{)}] The slit width is comparable to the angular size of the
galaxy. The  observed wavelength shifts are thus  averages of velocity
components coming from many parts of the galaxy.

\item[{(}3{)}] The galaxy is not seen edge-on, but with an inclination
angle $i$. Radial  velocities have thus to be  divided by $\sin{i}$ to
be transformed into rotation velocities.

\item [{(}4{)}] The radial  velocity curves are smeared by convolution
with the seeing profile, both in spatial and spectral directions.
\end{enumerate}

The procedure  we use assumes a  simple analytical mass  model for the
host galaxy, which includes a thin disk, a spherical dark matter halo,
and a central point mass. The velocity at a distance $r$ away from the
galaxy's center is then given by:\\
 
\begin{equation}
\centering
V_{\mathrm{mod}}(r) = \sqrt{-r(F_{\mathrm{disk}}+F_{\mathrm{dark}}+F_{\mathrm{CM}})}
\label{vmod}
\end{equation}

The radial forces in equation (\ref{vmod}) are computed using a
Kuzmin potential for the disk:
\begin{equation}
 F_{\mathrm{disk}}(r,z) = - \frac{r^2 G M_{\mathrm{disk}}}{(r^2+(a+|z|)^2)^\frac{3}{2}} 
\end{equation}
where we assume $z=0$, hence neglecting the thickness of the
disk. $M_{\mathrm{disk}}$ is the total mass of the disk and $a$ measures its
scale length.

\begin{figure}[t]
\centering                
\includegraphics[width=6cm]{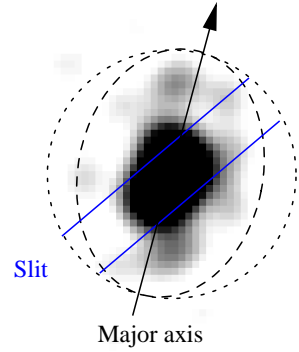}
\includegraphics[width=5cm]{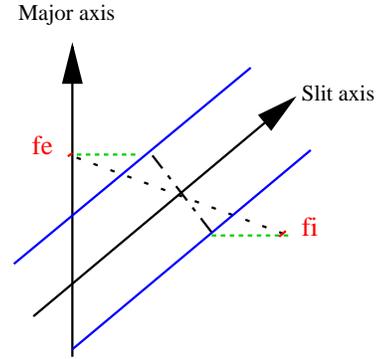}
\caption{Schematic illustration of  geometrical effects considered for
deriving the velocity curves. {\it  Top }: the original face-on galaxy
is assumed to be circular (dotted line). Seen with an inclination $i$,
it  appears elliptical  (dashed line).  In addition,  the slit  is not
aligned with the major axis of the inclined galaxy. {\it Bottom }: The
mean  velocity  at  one  position  along  the  slit,  which  would  be
integrated along a line perpendicular to the slit (dashed-dotted line)
if the galaxy was seen face-on,  has now to be integrated along a line
which is  inclined (and lengthened)  with respect to the  slit (dotted
line; see text).}
\label{schemavmoy} 
\end{figure}

Dark matter is assumed to be spherically symmetrical about the
galaxy's center. The force it creates at the distance $r$ away from
the center, given a radially dependent density distribution is:
\begin{equation}
F_{\mathrm{dark}}(r) = - \frac{4\pi G \rho_0 r_0 ^3}{r}(\frac{r}{r_0}-\arctan{\frac{r}{r_0}})
\end{equation}
where $r_0$ is the scale length for the mass distribution within the
dark matter halo and $\rho_0$ its central density (Binney \& Tremaine 
\cite{binney}).

The force from the central mass $M_{\mathrm{cm}}$ is simply 
\begin{equation}
 F_{\mathrm{CM}}(r) = - \frac{G M_{\mathrm{cm}}}{r}
\label{forceCM}
\end{equation}

The  five  parameters $M_{\mathrm{disk}}$,  $a$,  $\rho_0$, $r_0$  and
$M_{\mathrm{cm}}$ are determined  by a least squares fit  to the data,
i.e., to the measured velocity  shifts at each spatial pixel along the
slit. We therefore minimize the following $\chi^2$:
\begin{equation}
\chi^2=\sum_{j=1}^{N}\frac{1}{\sigma_{j}^2}                  
\Large{\{}
(s*V_{\mathrm{mean}})_{j}\ \sin{i} - d_j\Large{\}}^2
\label{chicarre}
\end{equation}
where $N$ is the total number of pixels along the slit, and where $d_j
\pm \sigma_j$ is the measured radial velocity shift at pixel $j$ along
the  slit.  $V_{\mathrm{mean}}$  is  a  one  dimensional  vector  that
contains  the  mean  (see  next  paragraph for  the  signification  of
``mean'') velocity predicted by our  mass model at pixel $j$ along the
slit. Due to the seeing, this  velocity is affected by blurring in the
spatial  direction,  hence the  need  for  the  convolution by  the  1
dimensional PSF $s$ obtained from the spectrum of stars. As the galaxy
has an inclination  $i$, we multiply the result  of the convolution by
$sin~i$ before comparing with  the observed velocity $d_j$.  The error
$\sigma_j$  on  the  radial  velocity  measurements  is  estimated  by
comparing the measured $d_j$ for  the same emission lines in different
grisms. This was possible for the H$\alpha$, [N\ion{II}], H$\beta$ and
[O\ion{III}] emission lines.

All the calculations  in Eq. \ref{chicarre} assume that  the galaxy is
seen  face-on,  the  inclination  being  taken  into  account  by  the
$\sin{i}$ term.  Due to the non-zero angle $\phi$ between the slit and
the major axis of the galaxy,  the velocity in Eq. \ref{chicarre} is a
weighted mean $V_{\mathrm{mean}}$ of several velocity components 
across the slit.

\begin{figure}[t]
\centering                
\includegraphics[width=6.cm]{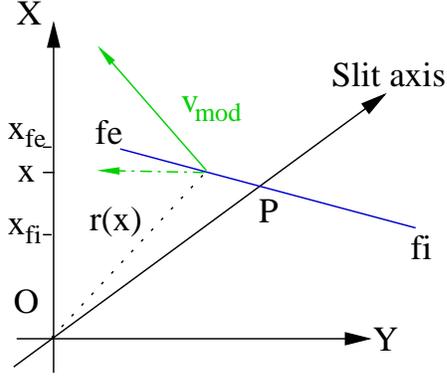}
\caption{Description of the integration  domain for the computation of
$V_{\mathrm{mean}}(P)$. The major axis of the galaxy is along X. $f_e$
and  $f_i$  are  the  limits  of  the  integral  also  shown  in  Fig.
\ref{schemavmoy}.   Only   the  component   of   the  model   velocity
$V_{\mathrm{mod}}$ along  the Y axis contributes to  the integral when
the galaxy is seen inclined.}
\label{vmeanfig} 
\end{figure}

For each  pixel $j$ along  the slit, lets  say at point $P$  (see Fig.
\ref{vmeanfig}), one can compute a mean velocity as follows:

\begin{equation}
\centering
V_{\mathrm{mean}}(P)=\frac{1}{I_{\mathrm{tot}}} \int_{X_{\mathrm{f_{i}}}} ^{X_{\mathrm{f_{e}}}} I[r(x)] \cdot  V_{\mathrm{mod}}[r(x)] 
\cdot \frac{x}{r(x)} \cdot dx
\label{vmean}
\end{equation}

$V_{\mathrm{mean}}(P)$ is  the average of all  the velocity components
located along  the dash-dotted line in Fig.  \ref{schemavmoy}.  If the
galaxy   was  seen   face-on  (as   we  would   like  it   to   be  in
Eq. \ref{chicarre}),  this line would  be rotated and  lengthened, and
mapped onto the dotted line of Fig. \ref{schemavmoy}.  The integral in
Eq.   \ref{vmean}  runs   along  this   dotted  line. Its  equation
(Eq.   \ref{equation_P})   is   easily   obtained   from   geometrical
considerations, given  the inclination $i$ of the  galaxy with respect
to the plane  of the sky and given the orientation  $\phi$ of the slit
relative to the major axis of the galaxy:

\begin{equation}
y= \frac{\cos{\phi}}{\cos{i}} \ (x - \frac{x_{\mathrm{P}}}{\cos{\phi}}).
\label{equation_P}
\end{equation}

$x_{\mathrm{P}}$ is the coordinate of the point $P$ on the major axis $X$
(Fig. \ref{vmeanfig}) of the galaxy.

The integration  in Eq. \ref{vmean}  is then performed as  follow: for
each position x  along the major axis of the  galaxy, we calculate the
distance  $r(x)$  from the  galaxy's  center O  to  the  point of  the
integration line that  is considered. This distance is  used to weight
each  velocity  component  $V_{\mathrm{mod}}[r(x)]$  of  the  velocity
model,  according to  the  light  distribution of  the  galaxy in  the
emission line. We have estimated  the light distribution $I(r)$ of the
galaxy, assuming it  is intrinsically symmetrical about  the center of the
galaxy. We  model it  by fitting a  Moffat profile to  the 1-dimensional
intensity profile  along the emission line,  and applying deprojection
(note that the fit of  the emission line is one-dimensional, but since
$I(r)$ is known  analytically, one can compute weights  in 2 dimensions
when  calculating  the  integral  in  Eq.  \ref{vmean}).  When 
doing the fit, the Moffat profile is convolved with the PSF. 
Correct normalisation  is
assured  by   $I_{\mathrm{tot}}$:  the  total  flux   of  the  regions
contributing to the measured emission at position $P$ along the slit.

The term $\frac{x}{r(x)}$ in Eq.  \ref{vmean} is the projection of the
rotation velocity  on the  minor axis of  the galaxy, which  will give
rise to a radial velocity once inclination is taken into account.  The
component of the velocity along  the X-axis does not contribute to the
mean velocity calculation, once inclination has been taken into 
account.
 
The intersections between the integration line and the slit edges, $f_i$
and $f_e$, give us the limits of the integration, as shown in 
Figs. \ref{schemavmoy} and \ref{vmeanfig}).


 
\begin{figure}[t]
\centering
\includegraphics[width=8cm]{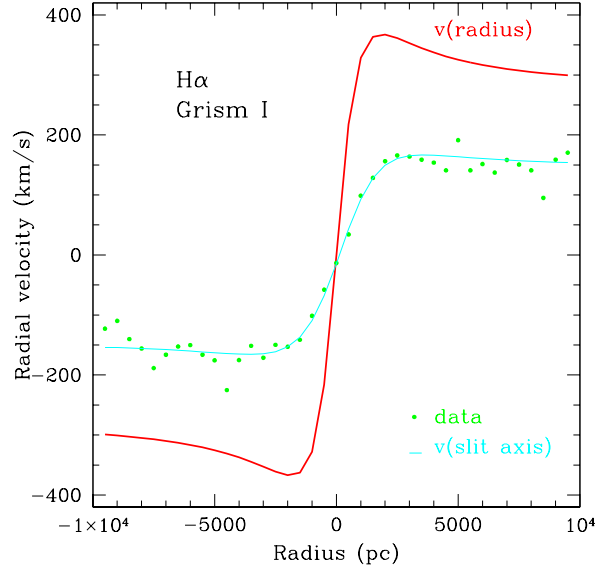}
\caption{Inclination/seeing   corrected   velocity   curves  for   the
H${\alpha}$ and  [NII] emission line  ($I$-grism only): the  dots show
the velocity measurements.  The solid line superposed is the fit of our
three components model after  taking into account misalignment between
the slit and the major axis  of the galaxy (see text). This solid line
shows the  variation of $V_{\mathrm{mean}}$ {\it along  the slit}.  It
is  then   corrected  for  inclination  and   seeing,  following  Eq.
\ref{chicarre}, to give the true velocity as a function of distance to
the center of the galaxy.}
\label{fitexpl}
\end{figure}

An example  of a inclination/seeing-corrected velocity  curve is given
in  Fig. \ref{fitexpl}  for the  H$\alpha$/N[\ion{II}] lines.   In our
calculations of the linear scales  and distances we have used $H_0$=65
km.s$^{-1}$Mpc$^{-1}$, $\Lambda=0$,  and $\Omega_M=0.3$, leading  to a
scale of $1\arcsec = 2.5$ kpc.

\begin{figure}[t]     
\centering          
\includegraphics[width=8cm,height=8cm]{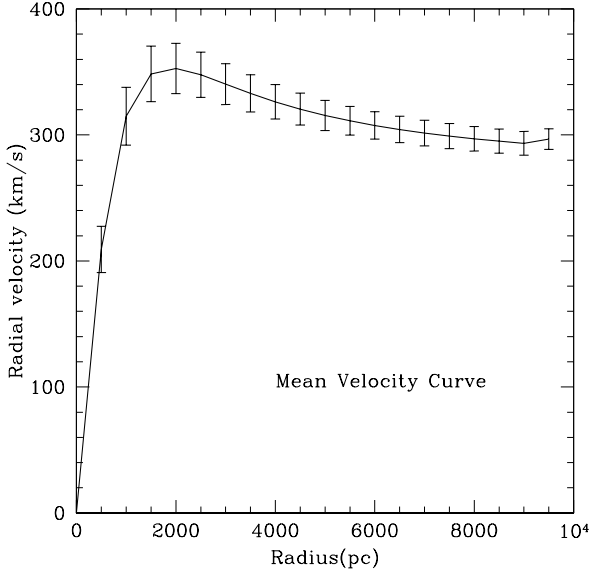}
\caption{Mean velocity curve
between four models fitted on the H${\alpha}$ line
(2 grisms) and on the  H${\beta}$ line (2 grisms). The error bars
are derived from the dispersion between the four curves. The result is
corrected for seeing and slit effects, as well as for inclination.}
\label{velcurv} 
\end{figure}

\subsection{Velocity curves: results }

We give in Table \ref{param} the best fit  for the 5 parameters  
of our mass model.  The mean velocity curve, as  extracted  from 
the H${\alpha}$  and H${\beta}$  lines is  displayed    in
Fig.  \ref{velcurv}. This mean corresponds to four independent curves
obtained for the  H${\alpha}$ and the H$\beta$ lines, both seen
in two different grisms. 

Table \ref{masses} gives our mass estimates which are weighted means of 
the four fits presented in Table \ref{param}, the weights being a function 
of the residuals after modeling of the data obtained in each 
individual grism. These weights are the same as the ones applied to
compute the mean velocity curve of Fig. \ref{velcurv}.  

We investigate in Table \ref{errorbar} the
effect of the error on the inclination of the galaxy and on its PA. 
Uncertainties on inclination  and orientation of the major axis
lead to  error bars on masses, which  are summarized in
Table \ref{errorbar}.

\begin{table}[t]
\caption{Mass model parameters for our velocity curves in each
individual grism and for each emission line.  $a$ and $r_0$
are  in   parsec,  $\rho_0$   in   M$_{\odot}/$pc$^3$  and
$M_{\mathrm{disk}}$ and $M_{\mathrm{cm}}$ in M$_{\odot}$.}
\begin{tabular}{lcccccc} 
\hline
Line & Grism & $a$  & $M_{\mathrm{disk}}$   & $\rho_0$ & $r_0$ & $M_{\mathrm{cm}}$ \\ 
\hline\hline
H${\alpha}$ & R &  1088 &  7.9 $10^{10}$&  0.30& 2033&  6.1 $10^7$
\\
H${\alpha}$ & I &1163 &  8.6  $10^{10}$ & 0.31 & 2074&  4.8 $10^7$
 \\
H${\beta}$ & B &908 & $4.8 \ 10^{10}$ & 0.56 & 1753 & 3.5 $10^6$
  \\
H${\beta}$ & R & 1428 &  5.9 $10^{10}$ & 0.23 &  2196 &  2.5 $10^5$ 
 \\
\hline
\end{tabular}
\label{param}
\end{table}

\begin{table}[t]
\caption{Weighted means of the masses obtained in Table \ref{param} 
where the weights are computed from the residuals of each individual
fit. M$_{\mathrm{tot}}$ is the sum of M$_{\mathrm{disk}}$ and 
M$_{\mathrm{dark}}$.  Units are solar masses.}
\begin{tabular}{lllll} 
\hline
Masses  &  $r< 1$ kpc&$r< 3$ kpc& $r< 10$ kpc\\
\hline\hline
M$_{\mathrm{disk}}$ &  1.9$\pm 0.3 \ 10^{10}$&5.1$  \pm0.7\ 10^{10}$ &   7.0$  \pm0.9\ 10^{10}$
\\
M$_{\mathrm{dark}}$ &  1.1$  \pm 0.3 \ 10^{9}$ & 1.6$  \pm0.3\ 10^{10}$ &  1.2$  \pm0.1\ 10^{11}$
 \\
\hline
\\
M$_{\mathrm{tot}}$ & 2.0$ \pm 0.3 \ 10^{10}$ & 6.7$  \pm1.0\  10^{10}$ &  1.9$  \pm0.2\ 10^{11}$ \\
\hline
\end{tabular}
\label{masses}
\end{table}

\begin{table}[t]
\caption{Effect of the error on  the inclination ($32 \pm 2 ^{\circ}$)
and PA ($38 \pm 3 ^{\circ}$) of the host galaxy. Results are given for
weighted means of 4 velocity curves, as in Table \ref{masses}, varying
the inclination and the PA within their error bars.}
\begin{tabular}{lllll} 
\hline
Masses (M$_{\odot}$) &  $r< 1$ kpc&$r< 3$ kpc& $r< 10$ kpc\\
\hline\hline
Inclination& & & \\
M$_{\mathrm{disk}}$ & $ 1.9 _{+0.2}^{-0.2} \ 10^{10}$& $5.1 _{+0.6}^{-0.6}\  10^{10}$ &   $7.0 _{+0.8}^{-0.8}\  10^{10}$
\vspace*{1mm}
\\
M$_{\mathrm{dark}}$ &$  1.1 ^{-0.02}_{+0.2}\ 10^{9}$ & $1.6  ^{-0.1}_{+0.2}\ 10^{10}$ &  $1.2  ^{-0.1}_{+0.1}\ 10^{11}$
 \\
\hline
\\
M$_{\mathrm{tot}}$ & $ 2.0 ^{-0.2}_{+0.2} \ 10^{10}$& $6.7 _{+0.8}^{-0.7}\ 10^{10}$ & $ 1.9 _{+0.2}^{-0.2}\ 10^{11}$ \\
\hline
\hline
Major axis&  & & \\
M$_{\mathrm{disk}}$ &  1.9$ ^{+0.02}_{-0.03} \ 10^{10}$&$5.1 ^{+0.1}_{-0.2}\ 10^{10}$ &   $7.0 ^{+0.2}_{-0.2}\ 10^{10}$
\vspace*{1mm}
\\
M$_{\mathrm{dark}}$ & $ 1.1 ^{+0.1}_{+0.04}\ 10^{9}$ &$ 1.6 ^{+0.1}_{-0.01}\ 10^{10}$ & $ 1.2 ^{+0.01}_{-0.05} \ 10^{11}$
 \\
\hline
\\
M$_{\mathrm{tot}}$ & $ 2.0 ^{+0.03}_{-0.03} \ 10^{10}$& $6.7 ^{+0.2}_{-0.2}\ 10^{10}$ & $ 1.9 ^{+0.03}_{-0.07}\ 10^{11}$ \\
\hline
\end{tabular}
\label{errorbar}
\end{table}

These results  should be interpreted with  care, as there  is a strong
correlation between  the model parameters and since  the adopted model
is  simpler than  reality.  However,  while the  details  of the  mass
distribution  within  the  galaxy  remain  inaccurate,  we  can  still
determine the  total mass  of dark  matter and as  well the  disk mass
enclosed  in  a given  radius.  

Given   the   results   in   Tables  \ref{param},   \ref{masses}   and
\ref{errorbar}, we  infer a mass estimate of  M(r $<$ 1  kpc) =
$(2.0  \pm 0.3)  10^{10}$  M$_{\odot}$. The  corresponding mass,  
integrated over 10  kpc is M(r  $<$ 10 kpc) = $(1.9 \pm 0.3) 10^{11}$ 
M$_{\odot}$. Table \ref{errorbar} shows that an additional error 
of 10\% should be added, due in part to the uncertainty on the PA of the 
major axis of the host, but mainly to the uncertainty on its inclination.

\section{Stellar and Gas Content}

One of the primary goals of our study is to infer the characteristics
of the stellar population of quasar host galaxies: does the stellar
population of \obj\, match that of a normal galaxy ? Where does it lie
in the Hubble sequence ?

Let us first note that $H$ = 14.57 mag corresponds, with the same
cosmological parameters used earlier, to an absolute magnitude of
M$_H$ =$ -24.5$ mag. This, in addition to the profile fitting described 
in the previous sections, indicates a normal early type local spiral galaxy
(Bothun et al. 1985; Arimoto \& Jablonka 1991). While the
photometric data alone already allow to point out a plausible
morphological type for the host of \obj, spectroscopy is the key to
the determination of its stellar content. A few considerations have to
be made before looking at the spectral features in any detail.

First, galactic extinction has to be taken into account. The
extinction towards \obj\, is A$_B$ = 0.217 corresponding to
E(B-V)=0.05 (Schlegel et al 1998).  Our spectrum of \obj\, has been
corrected accordingly.

\begin{figure}[t]
\centering
\includegraphics[width=8cm]{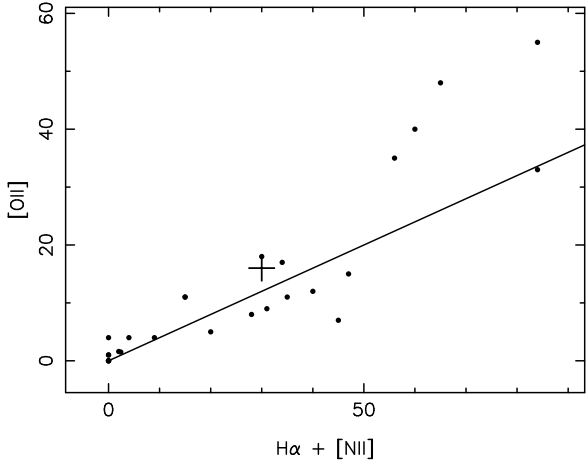}
\caption{The relation between the equivalent widths of 
H$\alpha$+[N\ion{II}]
and [O\ion{II}]. The black dots show the {\it normal} spiral galaxies
of Kennicutt's sample. The plain line correspond to [O\ion{II}] = 0.4
H$\alpha$+[N\ion{II}], mean relation of the galaxies dominated by
photoionization. The cross indicates the position of \obj.}
\label{HaOII}
\end{figure} 

Second, although no velocity standard stars were taken during the
observations, the resolution in the $R$-grism (4.9\AA; 3.6\AA\, at
rest-frame) is good enough to make use of the stellar library of Jacoby
et al. (1984) (FWHM =3.5 \AA). This gives a velocity dispersion of $\sim$ 170
km/s, which we only take as an indicative upper limit value.

Third, we observed \obj\, with three different grisms, each having a
different dispersion. The three spectra were combined in order to
cover the wide wavelength range 3670\AA\, to 9165\AA\, (3150\AA\, to
8073\AA\, once corrected from the redshift z=0.1355, or v=37837.4
km/s). In doing this, we have degraded the observations to the
resolution of the less resolving grism, i.e., 5.7 \AA\, (or $\sim$ 5
\AA\, at rest-frame).  

Finally, due to the low signal-to-noise ratio of the spectrum in its
bluer part, we consider only features from 4000 \AA\, on (3523\AA\,
at rest-frame).

We have only considered the integrated spectrum of \obj.  In fact, the
signal-to-noise of the measurements on each individual spatial resolution
element of the 2D data varies between 3 and 6, and therefore does not
give much hope in tracing any radial variation in the stellar
population, unless deeper data are obtained.

In the following, we measure well known indicators for the underlying
stellar populations, for the host of \obj, and we compare them with
the values obtained for two control samples: the one of Trager et al.
(1998) for the elliptical galaxies, and the one of Kennicutt (1992a,b)
for the spiral galaxies.

\begin{figure}[t]
\centering
\includegraphics[width=8cm]{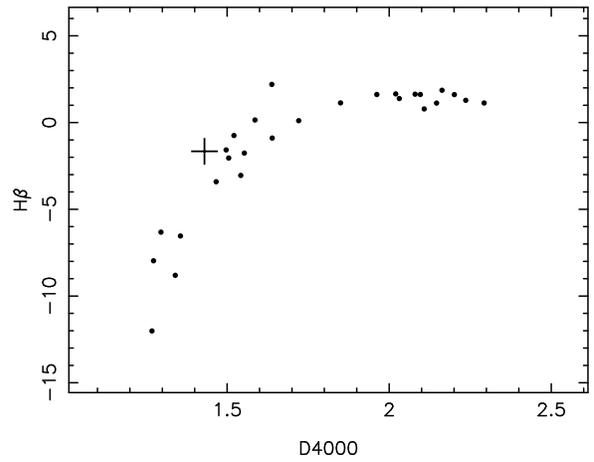}
\caption{The relation between the 4000\AA\, break and the equivalent
width of H$\beta$ for normal spiral galaxies (Kennicutt 1992) (dots)
and \obj\, (cross).}
\label{D4000Hbeta}
\end{figure} 

\subsection{Emission lines}

One of our concern was to verify that the deconvolution procedure had
indeed left out any non thermal excitation process.

We  have measured  the equivalent  widths (EW)  of the  most prominent
emission  lines  seen in  the  spectrum  of  \obj, i.e.,  [O\ion{II}],
[O\ion{III}](4959\AA), [O\ion{III}](5007\AA) and H$\alpha$+[N\ion{II}]
(see  Table \ref{eqw}).  

Kennicutt (1992a) provided measurement of the Oxygen lines [O\ion{II}]
(3727\AA), [O\ion{III}] (5007\AA) and H$\alpha$+[N\ion{II}]. He noted
that among galaxies with normal star formation rates
(EW(H$\alpha$+[N\ion{II}]) $\leq$ 40\AA), the [O\ion{III}] lines are
rarely detected at all (EW $\leq$ 0.5\AA).  For \obj, while
EW(H$\alpha$+[N\ion{II}]) $\sim$ 30\AA, we detect [O\ion{III}], with
an equivalent width of $\sim$ 4\AA.  This is the only and discreet
sign of an excitation process in or surrounding \obj\, which is not
directly related to star formation and, hence, which might be
triggered by the central AGN radiation field.  Kennicutt (1992a) also
reported that galaxies dominated by stellar photoionization follow a
mean relation EW([O\ion{II}]) $\sim$ 0.4
EW(H$\alpha$+[N\ion{II}])(with an rms of 50\% ).  \obj\, falls nearly
exactly on this relation, as shown in Fig. \ref{HaOII}.  Kennicutt's
galaxies are represented by plain dots, the mean relationship between
H$\alpha$+[N\ion{II}] and [O\ion{II}] is the solid line. The position
of \obj\, in this plane is indicated by a cross.

\obj\, exhibits a weak [O\ion{III}]  (5007\AA) line. 
We have left out from the Kennicutt sample the objects which displays
strong [O\ion{III}] (5007\AA) lines, when analysing in the next
section the absorption line features.  This ensures that we compare
\obj\, with objects of similar properties.

\begin{table*}[t]
\caption[]{Lick indices for the spectral features in the host galaxy
of \obj.}
\begin{tabular}{lcccccccc} 
\hline
Name  & D4000 & G4300 & Ca4455 & Mg$_2$ & Fe5270  & Fe5335 & [O\ion{III}] & H$\alpha$+ [N\ion{II}] \\
\hline
Value & 1.48  & 3.41  & 2.03   & 0.157  & 1.89    & 1.12   &  4.2   &   30.   \\
Error & 0.04   & 0.16  & 0.12   & 0.002  & 0.08    & 0.11   &  0.3   &    2.   \\ 
\end{tabular}
\label{eqw}
\end{table*}

To further investigate this matter, we have measured the line
intensity ratios, [O\ion{III}]/ H$\beta$ (0.05), [O\ion{III}]/
H$\beta$ (0.54), [N\ion{II}]/ H$\alpha$ (-0.20).
Due to the deconvolution procedure, the ratios between Oxygen and
Balmer lines are upper limits. Indeed, the Balmer lines are probably
underestimated. When placed in the diagnostic diagrams of Sodr\'e and
Stasi\'nska (1999), the host galaxy of \obj\, falls perfectly on
the sequence drawn by the normal spiral galaxies.

To summarize, our analysis of the emission lines suggests that \obj\,
is a perfectly normal spiral galaxy, where the gas emission is
dominated by ionization by stars rather than by the central AGN.

\begin{figure}[t]
\centering
\includegraphics[width=8cm]{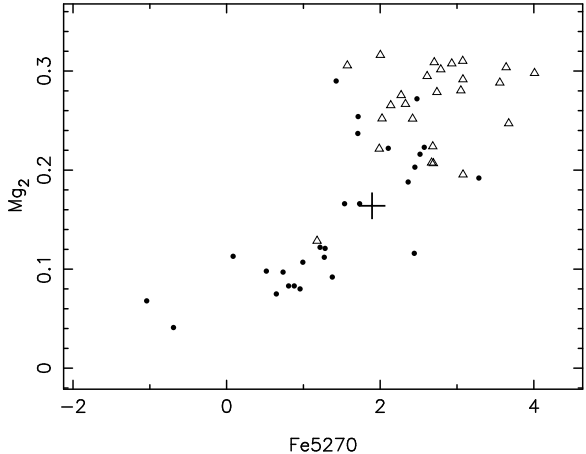}
\caption{The relation between the Fe5270 and Mg$_2$ index for the
normal spirals of Kennicutt et al. (1992) (dots) and the ellipticals
of Trager et al. (1998) (open triangles). As in Fig. \ref{HaOII}
and \ref{D4000Hbeta}, the cross indicates the position of \obj.}
\label{Fe5270Mg2}
\end{figure} 

\subsection{Absorption lines}

Our  spectra  have  sufficient  signal-to-noise to  measure  the  Lick
indices  for  several stellar  absorption  features.  As the  spectral
resolution  of the Lick  system is  lower than  the resolution  of our
data, we  have convolved the  spectrum of \obj\, with  the appropriate
Gaussian kernel, in order to match a 200 $\kms$ rest-frame resolution.

For a full compatibility between  the measurements done for \obj\, and
for  the  galaxy  control  sample  (see  previous  section),  we  have
re-measured the  Lick indices on  Kennicutt's galaxies in  exactly the
same way as for \obj.

A difficulty  when looking at the  stellar population in  \obj\, is to
avoid contamination due to the  gas emission. This emission can either
``fill-in''  absorption  lines --the  Balmer  lines  is  a well  known
example,  but Mgb  can be  affected  as well  (Goudfrooij \&  Emsellem
1996)-- or  contaminate   the  continuum  side  bands   used  for  the
measurements  of  the  equivalent  widths.  We  have  identified  four
absorption lines that are strong enough and free of this contamination
by emission lines. They correspond  to the Lick indices G4300, Ca4455,
Ca4668, Fe5270 and  Fe5335. With only a tiny  change in the definition
of  the continuum window  used for  the Mg$_2$  index, we  could avoid
contamination  by the  [O\ion{III}] (4959\AA)  emission line  for this
index. The small change we impose on the continuum side bands produces
a change of  at most $\pm$0.04mag (or 20\%) in  the measurement of the
Mg$_2$ index.  In addition to the indices for the absorption lines, we
also measure the  important calcium break at 4000\AA,  which gives the
D4000 index (Bruzual 1983).

Table  \ref{eqw}  summarizes  the  results of  our  measurements.  The
available   signal-to-noise  ratio  (per   \AA)  for   performing  the
measurements  varies between $\sim$  10 (blue  part) and  $\sim$25 (red
part).

The continuum shape  of \obj\, is rather flat,  indicative of on-going
star formation.  Similarly to the result of Fig. \ref{HaOII},
Fig. \ref{D4000Hbeta} puts
\obj\, among  normal spiral galaxies.  The latter figure  compares the
index D4000,  which measures the 4000\AA\, break,  with the equivalent
width  of  H$\beta$  which  is  indicative of  the  presence  of  star
formation. \obj\, nicely falls on  the sharp relation traced by normal
spirals.

As  for  the  chemical  abundances,   we  chose  to  present  in  Fig.
\ref{Fe5270Mg2} the Mg$_2$ and Fe5270  indices. They trace the
$\alpha$ and  iron peak-elements, respectively. Here  again, \obj\, is
located  exactly on  the Hubble  sequence, at  the location  of spiral
galaxies,  rather than  ellipticals.  Furthermore, the  values of  the
absorption  line   equivalent  widths  strongly   suggests the  stellar
population of an early type spiral galaxy.

\section{Conclusions}

We have undertaken a VLT  program aimed at unveiling the spectroscopic
properties  of quasar  host galaxies  for a  sample  of radio
quiet, bright quasars at low  redshift. Our long term 
scientific goal is double:
(1) to compare the  stellar populations of quasar hosts  with those of
other (``normal'') galaxies, and (2) to study their dynamics, as close
as possible to the central AGN.

The  present paper considers \obj\, as a test case.
An  important  part  of our  work  is  dedicated to  the
correction  of  geometrical effects  and  removal  of the  atmospheric
blurring. This  is crucial in the  central parts of  the galaxy, with
sizes comparable to the seeing disk.

We  find that  the stellar  population of  the host  galaxy  of \obj\,
compares  well with  that of  the  normal non-AGN  spiral galaxies  of
Kennicutt (1992a,b) and Trager  et al. (\cite{trager98}). They show no
trace of  enhanced star formation.   The interstellar medium  does not
show significant ionization by the central AGN. Its mass M(r $<$
10 kpc) = $(1.9 \pm 0.3) 10^{11}$ M$_{\odot}$ also compares
very well with that of  other spiral galaxies.  To summarize, the host
galaxy of \obj\, is a normal spiral galaxy.


\begin{acknowledgements}
The authors  would like to thank  Sandrine Sohy for help  with some of
the intensive programing involved in this work.  G\'eraldine Letawe is
a teaching assistant supported by the University of Li\`ege, (Belgium)
and is  partially funded by the  P\^ole d'Attraction Interuniversitaire,
P4/05  (SSTC, Belgium).   Fr\'ed\'eric Courbin  acknowledges financial
support  through  Marie  Curie  grant MCFI-2001-0242.   Chilean  grant
FONDECYT/3990024  and additional  support from  the  European Southern
Observatory  are  also  gratefully  acknowledged.   Dante  Minniti  is
supported in part by Fondap Center for Astrophysics.
\end{acknowledgements}

\end{document}